\documentclass[10pt]{article}
\usepackage{fancyhdr}
\usepackage{extramarks}
\usepackage{amsmath}
\usepackage{amsthm}
\usepackage[utf8]{inputenc}   
\usepackage[T1]{fontenc}
\usepackage{newunicodechar}   

\newunicodechar{α}{$\alpha$}
\newunicodechar{β}{$\beta$}
\newunicodechar{γ}{$\gamma$}
\newunicodechar{κ}{$\kappa$}
\newunicodechar{δ}{$\delta$}

\newunicodechar{–}{--}       
\newunicodechar{—}{---}      
\newunicodechar{’}{'}        
\newunicodechar{±}{$\pm$}

\newunicodechar{­}{}
\usepackage{amsfonts}
\usepackage{siunitx}
\usepackage{tikz}
\usepackage[plain]{algorithm}
\usepackage{algpseudocode}
\usepackage{multirow}
\usepackage{booktabs}
\usepackage{graphicx}
\usepackage{subfigure}
\usepackage[margin=1in]{geometry}
\usepackage{booktabs}
\usepackage{makecell}
\usepackage{array} 
\usepackage[colorlinks,linkcolor=black,anchorcolor=black,citecolor=black,urlcolor=blue]{hyperref}
\usepackage{hyphenat}
\usepackage{amsmath,bm}
\usepackage{booktabs}
\usepackage{mathtools}
\usepackage{amssymb}
\usepackage{tikz-cd}
\usepackage{caption}
\usepackage{capt-of}
\usepackage{mciteplus}
\usepackage{cite}
\usepackage{mathrsfs}
\usepackage[title,titletoc,toc]{appendix}
\usepackage{xr}
\usepackage{parskip}
\usepackage{soul}
\usepackage{textcomp}
\usepackage[colaction]{multicol}
\usepackage[switch]{lineno}
\usepackage{lipsum}
\usepackage{etoolbox}
\usepackage{longtable}
\usepackage{array}
\usepackage{tablefootnote}
\usepackage{ragged2e}
\usepackage{soul}
\newcolumntype{C}[1]{>{\centering\arraybackslash}p{#1}}
\usetikzlibrary{automata,positioning}
\usetikzlibrary{automata,positioning}

\usepackage{xr}
\begin{document}
	\title{Computational Drug Repurposing for Alzheimer's Disease via Sheaf Theoretic Population-Scale Analysis of snRNA-seq Data}
    
	\author{Sean Cottrell$^{1,2}$,  Seungmin Yoon$^3$, Xiaoqi Wei$^1$\footnote{Current address: Department of Mathematics, Center for Research in Scientific Computation, North
Carolina State University, Raleigh, NC 27695 USA}, Alex Dickson$^{2,4}$\footnote{
			Corresponding author.		Email: alexrd@egr.msu.edu}, and Guo-Wei Wei$^{1,4,5}$\footnote{
			Corresponding author.		Email: weig@msu.edu} \\
		\\
		$^1$ Department of Mathematics, \\
		Michigan State University, East Lansing, MI 48824, USA.\\
        $^2$ Department of Computational Mathematics, Science, and Engineering, \\
        Michigan State University, East Lansing, MI 48824, USA. \\
		$^3$ Department of Pharmacology and Toxicology, \\
        Michigan State University, East Lansing, MI 48824, USA. \\
        $^4$ Department of Biochemistry and Molecular Biology,\\
		Michigan State University, East Lansing, MI 48824, USA.  \\
		$^5$ Department of Electrical and Computer Engineering,\\
		Michigan State University, East Lansing, MI 48824, USA. \\
        \\
	}
	\date{\today} 
	
	\maketitle
	
	\begin{abstract}
	Single-cell and single-nucleus RNA sequencing (scRNA-seq  /snRNA-seq) are widely used to reveal heterogeneity in cells, showing a growing potential for precision and personalized medicine. Nonetheless, sustainable drug discovery must be based on a population-level understanding of molecular mechanisms, which calls for the population-scale analysis of scRNA-seq/snRNA-seq data. This work introduces a sequential target-drug selection model for drug repurposing against Alzheimer's Disease (AD) targets inferred from population-level snRNA-seq studies of AD progression in microglia cells as well as different cell types taken from an AD affected brain vascular tissue atlas, involving hundreds of thousands of nuclei from multi-patient and multi-regional studies. We utilize Persistent Sheaf Laplacians (PSL) to facilitate a Protein-Protein Interaction (PPI) analysis of AD targets inferred from differential gene expression (DEG), and then use a machine learning models to predict repurpose-able DrugBank compounds for molecular targeting. We screen the efficacy of different DrugBank small compounds and further examine their central nervous system (CNS)-relevant ADMET (Absorption, Distribution, Metabolism, Excretion, and Toxicity), resulting in a list of lead candidates for AD treatment. The list of significant genes establishes a target domain for effective machine learning based AD drug repurposing analysis of DrugBank small compounds to treat AD related molecular targets. 
    
	\end{abstract}
keywords: {Topological Data Analysis, Persistent Sheaf Laplacian, snRNAseq, Drug Discovery}
	
	\newpage
	
	{\setcounter{tocdepth}{4} \tableofcontents}
	\setcounter{page}{1}
 
	\newpage
	
	\section{Introduction}
 As of 2024, an estimated 6.9 million Americans over the age of 65 are living with Alzheimer's Disease (AD), and it consistently ranks among the top five most common causes of death for elderly individuals in the US. As life expectancy continues to rise, the incidence of age-related diseases also increases, and AD is projected to constitute a serious global health crisis by mid-century. Notably, Alzheimer's is the only top 10 cause of death in the United States with no known cure. Much like cancer development, Alzheimer's disease can be defined as a continuum of clinical and pathological events from normal aging to dementia- hitting primarily on neuro-inflammation, blood brain barrier disruption, and the accumulation of plaques. Detecting the pre-dementia stages of this continuum are notoriously difficult, especially so given the heterogeneity of AD- and so new diagnostic / treatment frameworks relying on the identification of biomarkers have been introduced and are gaining increasing attention in the medical community \cite{https://doi.org/10.1002/alz.14063}. 
 
 Single Cell and Single Nucleus RNA Sequencing have emerged in recent years as part of a new era in understanding disease pathology, drug discovery, and drug development. Single Cell RNA-seq is widely used to reveal heterogeneity in cells, which has given us insights into cell–cell communication, cell differentiation, and differential gene expression. Many tissues (e.g., the adult brain) are challenging to dissociate without altering transcriptional programs or losing fragile cell types. Single‑nucleus RNA‑seq (snRNA‑seq) complements single‑cell RNA‑seq to this end by profiling RNA from isolated nuclei rather than whole, live cells. Differential gene expression (DEG) analysis is a crucial tool in molecular biology and genetics. In particular, the analysis of interactomic networks constructed from these differentially expressed genes has had tremendous success in providing insights into a variety of biological and medicinal problems. Liang et al. applied Weighted Gene Coexpression Network Analysis (WGCNA) to the identification of key genes regulating Alzheimer's \cite{doi:10.3233/JAD-180400}. While notable, such methods are traditionally limited in handling only the low-dimensional, or pairwise, relations in the gene co-expression networks, which may result in an incomplete understanding of the complex, high-dimensional, non-linear nature of gene-gene interactions. 

 Du et al. proposed a novel topological differentiation technique to identify key genes from a protein–protein interaction network derived from DEGs relating to opioid and cocaine addiction \cite{dutop}. This topological differentiation technique serves as an example of the valuable perspectives offered by topological data analysis (TDA) to the study of biological networks through the application of persistent topological Laplacians\cite{wei2025persistent_2}. The first persistent topological Laplacian was introduced as a new generation of TDA models in 2019 \cite{wang2020persistent}. It has stimulated a variety of theoretical developments and led to remarkable applications \cite{su2025topological}. More recently, Wei and Wei have introduced the Persistent Sheaf Laplacian (PSL) for cellular sheaves, and described how to construct sheaves for a point cloud where each point is associated with a quantity that can be devised to embed some physical property or similarly, to emphasize a node in its network \cite{Wei2025persistent}. While the traditional Persistent Laplacian provides only a global topological description, the PSL is a local method, making it better suited to topological perturbations by analyzing the topology of the gene network with respect to a specific gene. In our context, raw differential expression magnitudes may emphasize genes that exhibit high fold-changes but do not significantly interact with other genes or connect important modules. Meanwhile, traditional topological Laplacians may emphasize genes, which are structurally significant but not significantly dysregulated, and therefore are simply noise or post‑transcriptionally regulated. A hub that barely changes at the RNA level likely does not drive pathology. Sheaf Laplacians address this shortcoming through their ability to simultaneously incorporate non-geometric information via cellular sheaves on the simplicial complex. Specifically, by labeling each node in the PPI complex by the magnitude of its dysregulation, we seamlessly unite both perspectives. 
 
 Additionally, previous studies utilizing the persistent Laplacian are limited by their analysis of only single sample data rather than full populational or atlas studies. Although scRNA-seq and omics data offer a revolutionary opportunity for patient-specific and personalized medicine, they do not translate directly to personalized drug development, as this is not sustainable on an industrial scale. Personalized medicine can be achieved with a specific selection or combination of drugs based on an individual's genetic biomarkers, but efficient drug development must be based on verifiable population-level molecular mechanisms. Therefore, it is necessary to identify commonly occurring targets on the population-level of scRNA-seq and/or spatial transcriptomic data for drug discovery. 
 
 Our goal is then to carry out a computational drug repurposing for AD targets identified via a population-scale snRNA-seq data analysis using the PSL. Our work then proceeds as follows. We obtained the GSE163577 dataset from the Gene Expression Omnibus database, providing single-cell profiling of the hippocampus and superior frontal cortex from 25 hippocampus and cortex samples across 17 control and 8 Alzheimer’s disease (AD) patients. These data were used to profile the major vascular and perivascular cell types of the human brain through 143,793 single-nucleus transcriptomes, enabling us to study cell-type–specific differential expression in the brain vasculature- a site of critical importance for the pathogenesis of neurodegeneration \cite{Yang2022}. In addition, we obtained 194,000 single nucleus microglial transcriptomes from CellBrowser. Each sample was segmented into various transcriptional states during AD progression from over 400 human subjects \cite{Sun2023}. This enables us to study the genetic heterogeneity in microglia cells during the temporal progression of neurodegeneration which drives neuro-inflammation trends as well as plaque clearance. In this study, we extracted a total of 50 potential gene targets. These topologically significant genes are then analyzed in the context of their surrounding PPI modules, providing coherent and targeted insight into pathology inducing pathways and machinery using gene set enrichment analysis. A small subset of the genes which corresponded to convergent disease inducing pathways across multiple lenses, and that have sufficient publicly available binding affinity data, are then targeted for computational drug repurposing efforts using DrugBank small compounds. The resulting compounds are further screened with Central Nervous System (CNS) related ADMET (Absorption, Distribution, Metabolism, Excretion, and Toxicity) analysis.   
 
\section{Results}
\begin{figure}[H]       
      \centering  
      \includegraphics[width=0.8\textwidth]{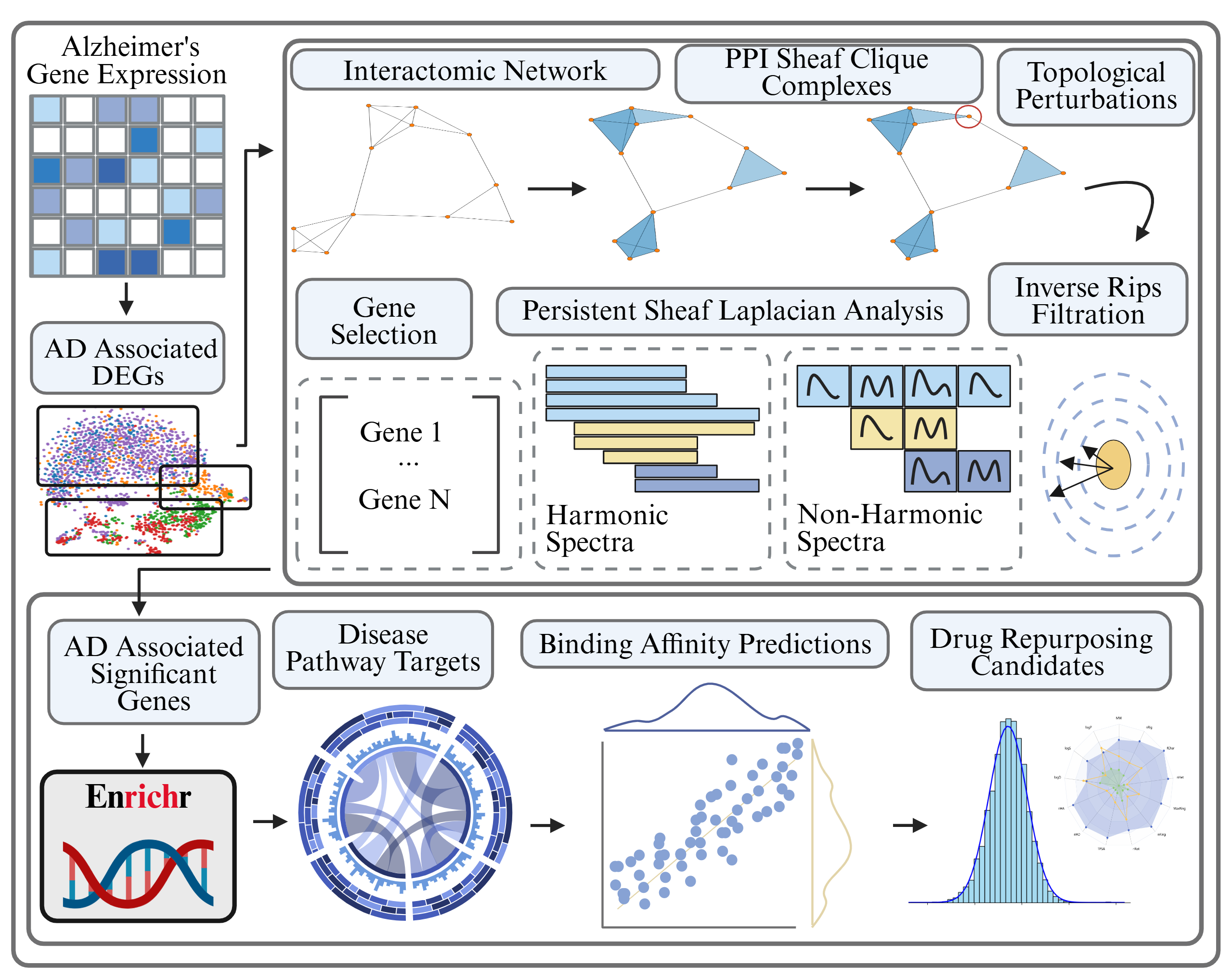}
      \caption{Schematic overview of our PSL analysis of PPI networks. Gene expression measurements are taken from various healthy and AD diseased cells and differentially expressed genes are extracted for different groups. These DEGs are used to construct PPI networks from the STRING database. The PPI networks undergo a clique expansion to construct a higher dimensional simplicial complex and each gene (0-simplex) is assigned a labeling reflecting its relative dysregulation magnitude in the group of interest. Gene specific topological perturbations and a filtration are then performed and the spectra of Persistent Sheaf Laplacians are analyzed to rank the genes by their significance. These significant genes are then targeted for further analysis via gene set enrichment and finally targeted drug repurposing. }
      \label{fig:flow}     
    \end{figure}

    In this section we walk through the results of the Persisent Sheaf Laplacian analysis of DEG derived PPI networks for cells taken either from a Human Brain Vascular Atlas or a Dynamic State Evolution Study of Microglia cells. The analysis was carried out by first obtaining a list of the top 200 differentially up-regulated genes within a set of cells of interest referenced against either control / homeostatic cells or cells sequenced earlier in the progression of neuroedegeneration. These DEGs were then used to construct cell type / state specific PPI networks and corresponding clique complexes, on which we could identify topologically significant genes with respect to each genes' respective magnitude of dysregulation via the cellular sheaf framework (Figure~\!\ref{fig:flow}). In this way, we theorize that pathology driving genes unite both topological significance with the non-geometric scale of their up-regulation. The top genes were chosen by intersecting the top 25 topological scores over each scale of filtration. Filtration steps were induced by varying the confidence of the PPI relations with thresholds of $[250,400,550,700]$, which ranges from a very loose certainty / strength of interaction to a strong, experimentally validated certainty. 

    \subsection{Disease Associated Expression Heterogeneity in Brain Vasculature}
    
\begin{figure}[H]       
      \centering  
      \includegraphics[width=\textwidth]{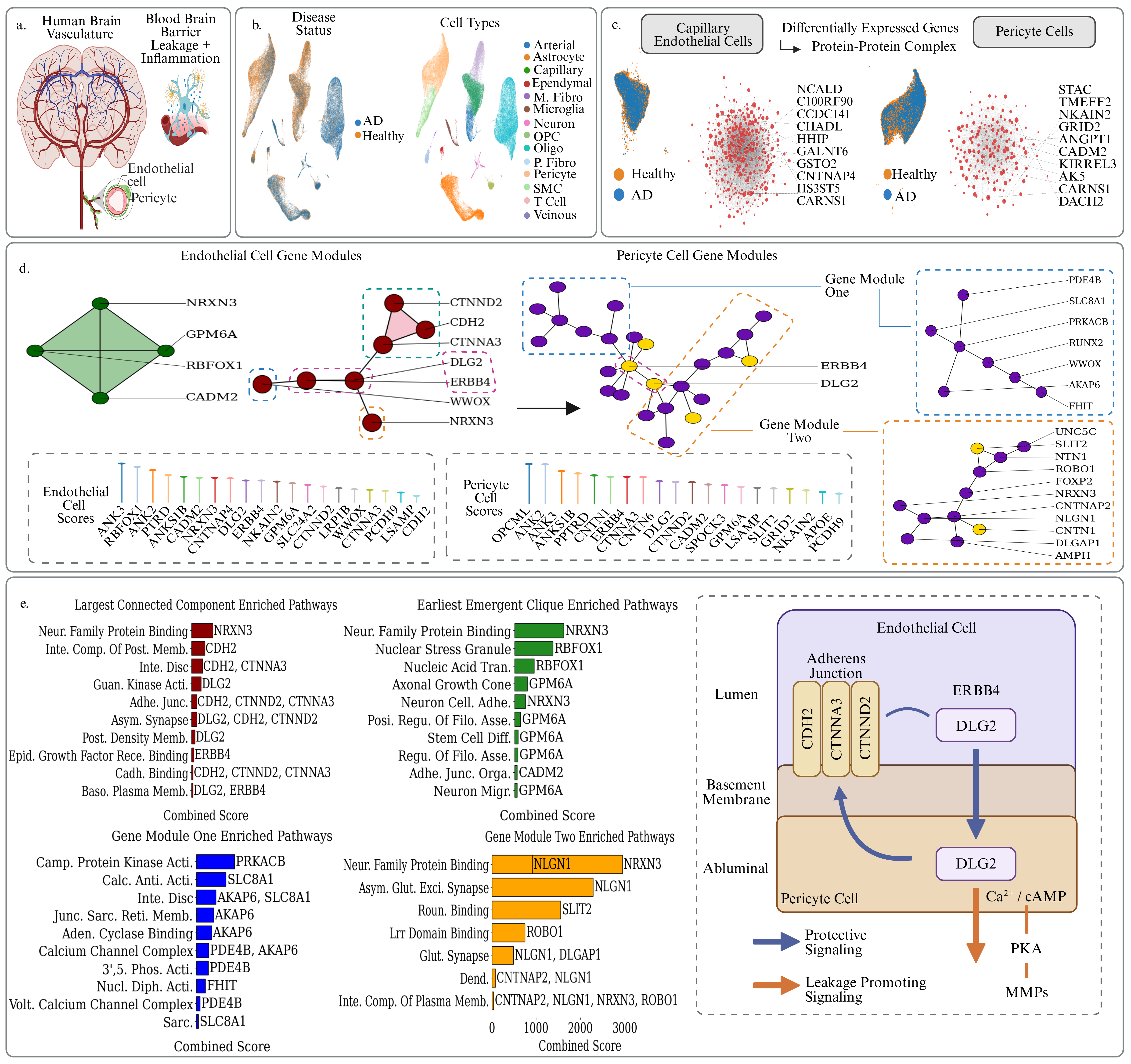}
      \caption{a. Depiction of the human brain vasculature structure and the place of capillary endothelial cells and perictye cells in that structure. These cells play a primary role in persistent BBB breakdown in AD which is a significant pathological driver. b. UMAP plots of the human brain vasculature atlas taken from Yang et. al. On the left cells are colored by their disease status and on the right cells are colored by their cell type. c. UMAP plots of diseased versus healthy endothelial and pericyte cells, as well as the respective PPI complexes of their up-regulated DEGs with the top genes in terms of log-fold changes labeled. d. Focused analysis of PPI structures surrounding our top scoring topologically significant genes from the PSL analysis. Cliques and connected components containing topological genes from endothelial and pericyte cells are extracted and their structure reveals a potential cross-compartment signaling cascade between protective adhesion modules and MMP inducing CA$^{2+}$ / cAMP / PKA machinery. e. The topological genes and their surrounding PPI structures were then used to conduct a pathway analysis via the GO databases in GSEApy, revealing coherent and coordinated pathways driving different aspects of BBB leakage pathology. A schematic diagram is also presented illustrating the proposed ERBB4-DLG2 mechanistic bridge.  }
      \label{fig:capperc}     
    \end{figure}

    Dysfunction of the brain vasculature is a known upstream driver, amplifier, and monitorable marker of Alzheimer’s pathology \cite{gullotta2023microglia} (Figure~\!\ref{fig:capperc}a). Alzheimer’s disease pathology generally begins years or decades before the onset of noticeable clinical symptoms, and one of its earliest hallmarks is dysfunction of the blood–brain barrier characterized by increased permeability, or leakage. Previous studies have indicated the involvement of several key signaling pathways, including inflammatory cytokine cascades, oxidative stress in endothelial cells, and pericyte dysfunction. These processes not only disrupt BBB integrity but also promote neuronal injury, driving neurodegeneration. Yang et. al developed a molecular map of the human brain vasculature by profiling the major vascular and perivascular cell types of the human brain through 143,793 single-nucleus transcriptomes taken from both diseased and control subjects \cite{Yang2022} (Figure~\!\ref{fig:capperc}b). Specifically, their study examined 17 post‑mortem individuals (9 affected by Alzheimer’s disease, 8 age‑matched subjects with no cognitive impairment). They looked at 25 fresh‑frozen tissue blocks: 17 hippocampus and 8 superior frontal cortex samples. They yieled 143,793 high‑quality nuclei captured with VINE‑seq after QC in total. By including both AD and healthy controls the atlas captures early and late vascular transcriptional changes linked to disease state, enabling a comprehensive DEG analysis in highly relevant cell types- endothelial capillary and Pericyte cells. Persistent Sheaf Laplacian analysis demonstrates an impressive ability to derive meaningful biological insights from this inter‑patient variation, which contrasts with previous Persistent Laplacian studies that were limited to single sample and more highly focused studies \cite{dutop}.

Endothelial cells line the brain’s capillaries and normally maintain barrier integrity.  Alzheimer’s-disease (AD) affected brains display altered gene-expression signatures in these cells, mirroring early blood–brain-barrier (BBB) leakage and neuro-inflammatory feed-back loops \cite{gullotta2023microglia} (Figure ~\ref{fig:capperc}a).  Early endothelial injury allows plasma proteins to enter the brain, accelerates microglial activation, and propagates synaptic loss and further damage to the BBB.  Pericyte cells, meanwhile, wrap the brain capillaries and instruct endothelial cells to adopt BBB programs. Our Persistent Sheaf Laplacian enabled differentiation tool allowed us to extract twenty topologically significant genes (Table \ref{tab:vascular_sig}) from the differentially expressed gene (DEG) set of both endothelial and pericyte cells for focused analysis.  These genes, together with their positions in the protein-interaction network and enriched pathways, are visualized in Figure ~\ref{fig:capperc}c.

\begin{table}[H]
\centering
\caption{Topologically significant genes in AD-affected vascular cells.  
E = Endothelial only; P = Pericyte only; B = Both cell types.}
\label{tab:vascular_sig}
\begin{tabular}{llll}
\hline
Gene & Cell type & log$_2$FC & p-value\\
\hline
ANK2      & B & 0.79,0.81 & 2.9e-04,6.4e-18 \\
ANK3      & B & 1.41,1.61 & 1.5e-05,1.7e-08 \\
ANKS1B    & B & 0.79,0.97 & 3.7e-02,3.7e-04 \\
APOE      & P & 0.72      & 1.5e-03 \\
CADM2     & B & 1.57,1.90 & 1.1e-37,1.8e-101 \\
CDH2      & E & 0.55      & 4.4e-20 \\
CNTN1     & P & 1.18      & 9.9e-57 \\
CNTN6     & P & 1.32      & 1.1e-20 \\
CNTNAP4   & E & 0.83      & 2.2e-09 \\
CTNNA3    & B & 1.28,1.40 & 2.3e-18,3.5e-55 \\
CTNND2    & B & 0.74,0.92 & 6.1e-13,1.4e-17 \\
DLG2      & B & 0.39,0.61 & 4.9e-11,2.6e-19 \\
ERBB4     & B & 0.54,0.66 & 2.6e-07,2.9e-07 \\
FHIT      & P & 0.36      & 1.3e-04 \\
GPM6A     & B & 0.85,0.97 & 1.2e-14,5.2e-20 \\
GRID2     & P & 1.78      & 1.1e-192 \\
LRP1B     & E & 0.43      & 2.7e-07 \\
LSAMP     & B & 1.22,1.38 & 1.9e-08,6.3e-43 \\
NKAIN2    & B & 1.09,1.17 & 2.9e-21,8.2e-34 \\
NRXN3     & B & 1.00,1.04 & 4.6e-07,9.6e-07 \\
OPCML     & P & 1.25      & 2.0e-15 \\
PCDH9     & B & 1.44,1.25 & 5.0e-82,8.6e-118 \\
PDE4B     & B & 1.13,0.41 & 9.5e-15,8.3e-08 \\
PRKACB    & P & 0.31      & 7.9e-03 \\
PTPRD     & B & 1.28,1.09 & 1.6e-09,3.4e-15 \\
RBFOX1    & E & 0.80      & 6.4e-02 \\
SLC24A2   & B & 1.68,1.63 & 3.8e-09,4.8e-12 \\
SLIT2     & P & 0.50      & 4.2e-04 \\
SPOCK3    & B & 0.97,1.13 & 3.3e-65,3.4e-06 \\
WWOX      & B & 0.33,0.58 & 5.4e-10,6.8e-52 \\
\hline
\end{tabular}
\end{table}

In AD-affected brains, the endothelial pathology is heterogeneous, so the DEG list is broad.  Examining how our significant genes assemble into sub-graphs and cliques allows us to yield more coherent pathway themes for analysis. For example, the functional relation between CDH2, CTNNA3, CTNND2, WWOX, DLG2, ERBB4, and NRXN3 is evident from their forming the largest connected component of significant genes in the PPI network (Figure \ref{fig:capperc}d). Pathway enrichment analysis via the GO Cellular Component and Molecular Function databases points to coordinated dysfunction of junctional integrity and signalling at the BBB, consistent with compensatory up-regulation during early damage (Figure ~\ref{fig:capperc}).  Several genes relate to cell–cell adhesion pathways: CDH2 and its catenin partners CTNNA3 and CTNND2 mediate tight contacts between endothelial cells and pericytes, preserving barrier integrity \cite{Kruse2019, Tan2023}.  Loss of this adhesion likely weakens the BBB as pericytes detach.  DLG2 carries a guanylate-kinase–like domain that provides a docking platform; MAGUK scaffolds organize adherens junctions and connect them to kinases such as ERBB4.  Because both DLG2 and ERBB4 are annotated to the basolateral plasma membrane and MAGUKs recruit polarity complexes, this interaction places ERBB4 on the abluminal face that meets the basement membrane and pericytes \cite{https://doi.org/10.1111/j.1742-4658.2012.08729.x, Castiglioni2022, Sorkin2008}. The relation therefore merits functional testing, as it touches both cadherin belts and basolateral polarity, which are essential in BBB integrity (Figure \ref{fig:capperc}e).

We next examined the pericytes which envelop the basement membrane.  We again identified twenty significant genes, nine of which overlap with the endothelial list, possibly suggesting a shared molecular backbone across the two vascular compartments (Table \ref{tab:vascular_sig}). It is interesting then to consider the context of these genes in the wider set of Pericyte DEGs. Extracting the PPI connected component which contains the greatest number of our significant genes yields the third sub-graph in Figure \ref{fig:capperc}. Here, the ERBB4 / DLG2 pair again appears, now linking a Ca\textsuperscript{2+}/cAMP-rich module (PDE4B, SLC8A1, AKAP6, FHIT, RUNX2, WWOX, PRKACB) with neuronal-adhesion genes (NRXN3, CNTNAP2, NLGN1, SLIT2/ROBO1, etc.).  This  may suggest an abluminal handshake in which the same scaffold couples adhesion–polarity machinery to leakage-promoting signaling on both sides of the basement membrane.  Together, the two cell-type significant gene subgraphs in Figure \ref{fig:capperc}d along with the corresponding enrichment analysis of these subgraphs in Figure \ref{fig:capperc}e seemingly converge on a model where an ERBB4-anchored MAGUK node links cadherin-belt repair in endothelial cells to chronic matrix-remodelling signals in pericyte cells via a PKA/Ca$^{2+}$/cAMP to MMP pipeline, potentially turning an initially protective instruction into a self-perpetuating leakage loop (Figure ~\ref{fig:capperc}e) \cite{Wang2017_2, Parker2019}. While some prior studies have reported barrier-protective effects of ERBB4 activation, such structure raises the alternative hypothesis that maladaptive, context-specific ERBB4 signaling may instead exacerbate chronic BBB permeability in AD. Further experimental work would be needed to validate these claims. 

The formation of cliques among the topologically significant genes in endothelial cells grants additional insight.  The earliest four-node clique to appear during simultaneous topological score and PPI edge filtration comprises NRXN3, CADM2, GPM6A, and RBFOX1- a module of neuronal-adhesion genes.  Brain capillary endothelial cells can express synaptic adhesion molecules \cite{Dilling2017, Yang2022}. It has previously been established that deletion of the Pcdh family member PCDH9 in microvascular endothelium increases paracellular permeability \cite{Gabbert2020}.  CNTNAP4, another synaptic adhesion molecule detected here, shows measurable endothelial expression \cite{Bottos2011}.  Such maladaptive expression may be the result of ambient noise, or rather, may indicate an aberrant attempt at barrier remodelling during leakage.  Most notably, in the pericyte analysis this neuronal-adhesion programme may couple to the Ca\textsuperscript{2+}/PKA/MMP axis through the DLG2 / ERBB4 bridge, potentially worsening leakage while repair genes remain chronically elevated.

Several significant genes also relate to cell stress and amyloid-beta handling.  LRP1B is a lipoprotein-receptor relative of LRP1, which ferries amyloid-beta out of the brain; its dysregulation may impair clearance \cite{Storck2015}.  PTPRD and WWOX are neurodegeneration-linked genes associated with stress signalling \cite{Liu2018, Chibnik2018}.  LSAMP has appeared in AD co-expression modules, though its mechanistic role is not fully understood.  Taken together, these findings underscore the popular notion of BBB disruption as a central pathological driver in AD and highlight the ERBB4 / DLG2 node as a promising, under-explored hypothesis for BBB leakage pathology.

\subsection{Temporally Driven Expression Heterogeneity in Disease Associated Microglia}
    The dynamic evolution of microglial states is a major driver of neuroinflammation and subsequent neurodegeneration as well as plaque clearance in AD. Sun et. al presented 194,000 single-nucleus microglial transcriptomes across 443 human subjects and diverse AD pathological phenotypes \cite{Sun2023}. They annotated a series of temporally developing cell states during AD progression, enabling researchers to study differential gene expression trends among different stages of neurodegeneration. The data also exhibits technical heterogeneity with libraries spanning 10x v2 and v3 chemistries, and has been batch corrected via Harmony. The ability for our PSL analysis to extract insights in spite of this technical heterogeneity is also in contrast to previous Persistent Laplacian based studies \cite{dutop}. This data represents one of the largest, modality‑matched censuses of human microglia cells to date, with enough subjects to dissect inter‑individual and stage‑specific expression programs. 

    \begin{figure}[H]       
      \centering  
      \includegraphics[width=\textwidth]{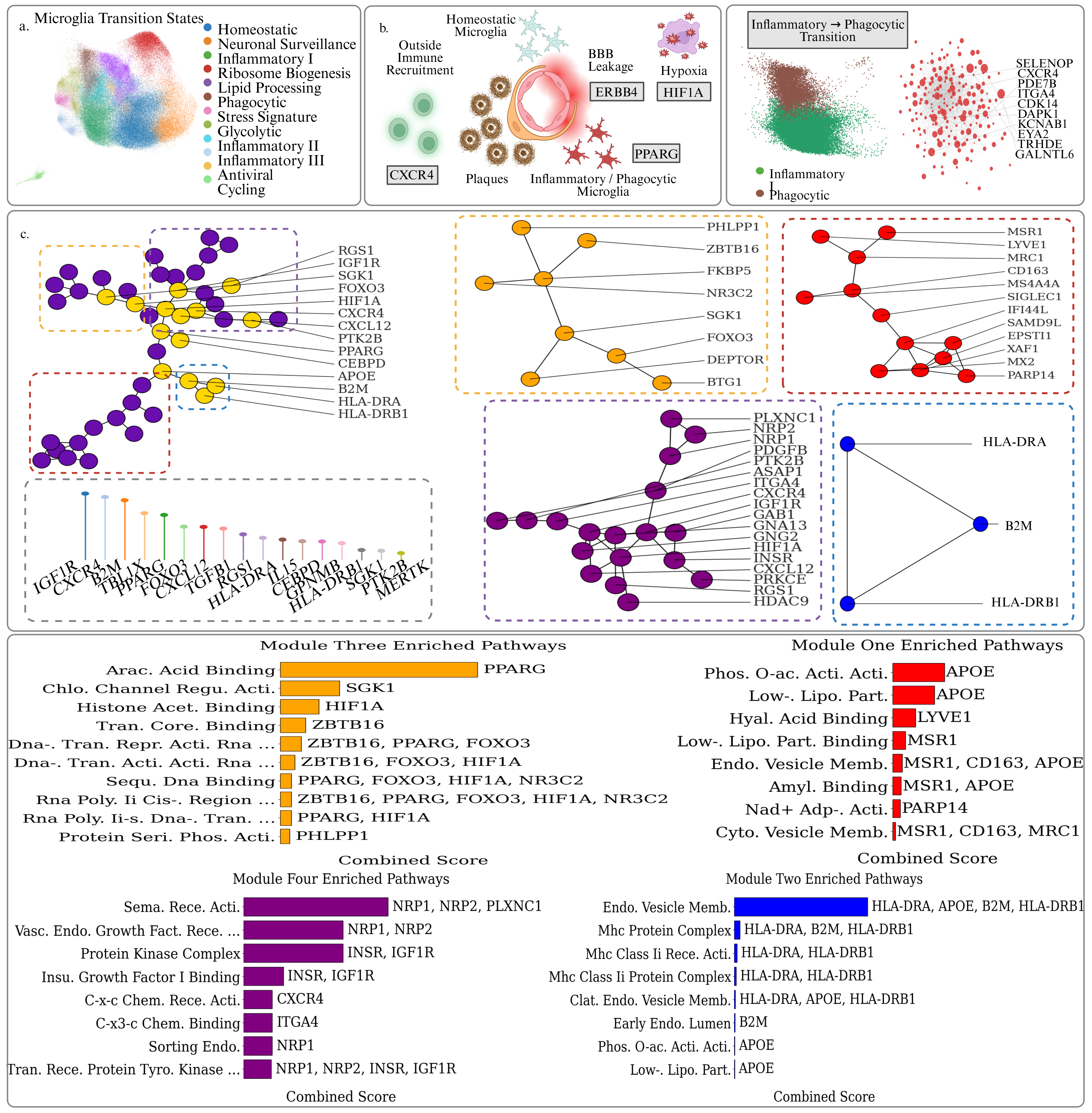}
      \caption{a. UMAP plot of the microglial state transitionary cells taken from Sun et. al. Cells are colored by their activation status. b. Schematic overview of the interactions between disease associated microglia cells and the BBB in AD. c. UMAP plots of the early Inflammatory and Phagocytic microglia cells, with up-regulated DEGs in the phagocytic state being used to construct a PPI complex. Top dysregulated genes in terms of logFC are labeled. d. Focused analysis of PPI structures surrounding our top scoring topologically significant genes from the PSL analysis. The largest connected component made up of topological genes is extracted and its structure reveals four primary axes joined by a central HIF1A / PPARG / APOE backbone. e. The topological genes and their surrounding PPI structures were then used to conduct a pathway analysis via the GO databases in GSEApy, revealing coherent and coordinated pathways driving different aspects of phagocytosis as well as hypoxia, antigen presentation, and outside immune recruitment. }
      \label{fig:microg}     
    \end{figure}
    
As AD progresses, microglia become increasingly triggered by signals from plaque buildup and enter a phagocytic stage. At this point the hallmark genetic profiling of microglia cells begins to reflect significant upregulation of pathways for debris clearance, antigen presentation, and immune recruitment. In Table \ref{tab:inflphag} we list 20 genes identified as being significant in our Persistent Sheaf Laplacian analysis of the PPI complex comprised of DEGs relative to an initial inflammatory state. These genes, along with their positions in the overall PPI network and topological significance scores, are shown in Figure~\ref{fig:microg}.

\begin{table}[H]
    \centering
    \caption{Significant genes in early-inflammatory vs.\ phagocytic microglia}
    \label{tab:inflphag}
    \begin{tabular}{lcc}
        \hline
        Gene & log$_2$FC & p-value \\
        \hline
        APOE                & 1.08 & $8.1\text{e-}165$ \\
        MERTK               & 0.93 & $2.9\text{e-}138$ \\
        FOXO3               & 0.96 & $2.3\text{e-}81$  \\
        CEBPD               & 1.15 & $5.4\text{e-}78$  \\
        RGS1                & 1.23 & $4.4\text{e-}59$  \\
        TGFBI               & 3.83 & $1.8\text{e-}53$  \\
        IL15                & 1.38 & $6.6\text{e-}42$  \\
        HLA-DRA  & 0.84 & $1.8\text{e-}40$ \\
        B2M                 & 0.48 & $2.3\text{e-}24$ \\
        PTK2B               & 0.73 & $5.4\text{e-}16$ \\
        CXCR4               & 4.18 & $1.1\text{e-}13$ \\
        DUSP1               & 0.87 & $1.5\text{e-}11$ \\
        GPNMB               & 3.77 & $2.4\text{e-}11$ \\
        HIF1A               & 0.23 & $1.8\text{e-}07$ \\
        PPARG               & 0.56 & $2.2\text{e-}07$ \\
        SGK1                & 0.49 & $1.9\text{e-}06$ \\
        TBL1X               & 0.68 & $7.9\text{e-}06$ \\
        IGF1R               & 0.63 & $1.4\text{e-}05$ \\
        HLA-DRB1 & 0.84 & $2.7\text{e-}05$ \\
        CXCL12              & 1.96 & $6.0\text{e-}05$ \\
        \hline
    \end{tabular}
\end{table}

We are interested in the positioning of these genes in the overall context of other up-regulated genes. If we extract the connected component containing the largest number of our significant genes, we obtain the network configuration in Figure \ref{fig:microg}. This resolves into four functional gene modules that converge on an HIF1A–PPARG–APOE backbone. All three genes are significantly up-regulated and achieve high mean topological scores across PSL filtrations (Table \ref{tab:inflphag}). The red module (MSR1, CD163, LYVE1, etc.) pathway analysis demonstrates enrichment of amyloid-$\beta$ binding and endocytic-vesicle membrane, indicating receptor-mediated tethering and phagocytosis of plaque material \cite{KoenigsknechtTalboo2004}. Plaque-associated hypoxia stabilises the HIF1A gene; its interaction with the p300/CBP complex and with Pol-II promoters is reflected in the orange enrichment terms histone acetyl-lysine binding and RNA-polymerase II cis-regulatory DNA binding \cite{Ogunshola2009, Ema1999}. The orange module itself—composed chiefly of PPARG, FOXO3, SGK1, ZBTB16, HIF1A and NR3C2—encodes the metabolic and transcriptional circuitry that equips microglia to digest the plaque cargo internalised by the red cluster. Its leading pathway, arachidonic-acid binding, singles out PPARG as a long-chain fatty-acid sensor that switches on lipid-oxidation and cholesterol-efflux programs \cite{Kliewer1997}. Chloride-channel regulatory activity is driven by SGK1, an Akt-responsive kinase that preserves ionic balance and prevents lysosomal rupture during intense lipid catabolism \cite{https://doi.org/10.1096/fj.12-218230}. The recurring chromatin terms—histone acetyl-lysine binding (HIF1A), transcription-core binding (ZBTB16) and multiple DNA-binding categories listing PPARG, FOXO3, ZBTB16—capture a coordinated transcriptional wave in which FOXO3 supplies antioxidant and autophagy genes, ZBTB16 remodels lipid-responsive promoters, and PPARG transactivates APOE \cite{Mammucari2007, Yue2009, Hu2022}. By linking hypoxia-stabilised HIF1A to PPARG-driven lipid detoxification—and by generating APOE particles that dock directly onto the red mocule's plaque-handling receptors—the orange module serves as the metabolic licence that allows microglia to clear A$\beta$ while attempting to avoid self-inflicted lipid and oxidative stress. Critically, the HIF1A–PPARG backbone is at the core of this interacting machinery. 

The blue module reflects the interplay between HLA-DRA, HLA-DRB1, and B2M. Their top enrichment terms—MHC-protein complex and clathrin-mediated endocytosis—mark the assembly and surface trafficking of both MHC-II and MHC-I peptide complexes. This configuration signals that microglia have progressed beyond mere phagocytosis and are now poised to additionally activate T cells \cite{Cash1993}. IL-15, while not included in this module, was identified as topologically significant and likely sits at the fringe of this module as it is a cytokine that supports survival of T cells. Microglia producing IL-15 could attract peripheral immune cells to enter the brain in AD, or activate resident T cells, linking innate and adaptive immunity in neurodegeneration \cite{Lee1996}. Similarly, CEBPD is a transcription factor that is upregulated by inflammatory stimuli and drives expression of cytokines and complement components. Its presence indicates a feed-forward loop reinforcing the microglial inflammatory phenotype and outside immune recruitment \cite{Ko2012}. RGS1, meanwhile, is known to modulate chemokine receptor signaling in immune cells, likely tuning the intensity of this antigen response \cite{Bowman1998}. The core of the purple gene module revolves around the chemokine pair CXCL12–CXCR4 and is highlighted in the pathway analysis for “C-X-C chemokine-receptor activity” and “VEGF signalling.” CXCL12 gradients, up-regulated in hypoxic plaque niches, guide CXCR4-expressing leukocytes—including T cells primed by IL-15—toward these lesions \cite{Arno2014}. Together, the two modules form a coordinated gateway: the purple arm orchestrates chemotactic recruitment, and the blue arm presents plaque-derived peptides to arriving T cells, seamlessly linking innate microglial activation with an adaptive immune response that can propagate neuroinflammation in Alzheimer’s disease.

    \subsection{Molecular Targeting Assessment}

    Having studied the topologically significant genes as well as their positioning in the larger PPI complex and its corresponding enriched pathways for each of our cell groups, we can arrive at some convergent conclusions with regards to meaningful hypotheses and therapeutic targets. In both sets of vascular cells we found genes involved in the breakdown of the Blood Brain Barrier, particularly due to loss of cell adhesion and endothelial-pericytes contact causing increased barrier permeability. Interestingly, examining the positional context of the topologically significant genes in both cells revealed the placement of ERBB4 at a putative cross-compartment hub. Its C-terminal PDZ-binding motif can dock onto MAGUK scaffolds such as DLG2, which in turn link adherens-junction complexes to intracellular kinases. We therefore hypothesise that ERBB4-DLG2 assemblies couple barrier-repair signalling in endothelial cells to Ca$^{2+}$/cAMP–PKA-driven MMP activity in pericytes, potentially converting an initially protective barrier maintenance response into machinery that drives chronic leakage. Targeted (either inhibitory or activating) small-molecule modulation of ERBB4 becomes an attractive line of investigation, especially if future experiments were to establish a truly causal role for this axis in BBB permeability. Importantly, this validates the potential for a PSL based analysis of biological networks to present novel and testable biological hypotheses. 

    From our analysis of inflammatory to phagocytic microglia transition, several more genes and pathways emerge as promising targets for drug repurposing. PPARG was identified alongside HIF1A as two of the central players uniting multiple pathology related machineries involving inflammation, plaque clearance, and hypoxia response. PPARG agonists have been tested in AD with some evidence that they can suppress microglial activation and also enhance the efficiency of A$\beta$ clearance \cite{Yamanaka2012}. In one of the PPI modules, the CXCR4-CXCL12 axis was positioned centrally, and pathway enrichment analysis revealed it as a primary driver of the inflammatory state caused by additional outside immune recruitment. CXCR4 antagonists might tamp down the deleterious pathogenic immune cell recruitment and microglial positioning. Finally, the central role of HIF1A in multiple modules highlights the importance of plaque induced hypoxia in AD affected micro-environments. It has been proposed that inhibition of Prolyl-hydroxylase (EGLN) is a potentially effective approach to enhancing the protective activity of HIF1A \cite{Lin2024}. Together, these targets represent a multi-pronged approach to reinforcing the BBB, quelling microglial and immune cell over-activation, boosting clearance of toxic proteins, and improving the brain's hypoxic response. Furthermore, this demonstrates the capacity for PSL based analysis of biological networks to fully recover the hallmark axes of disease associated microglia function from otherwise noisy differential expression data.  

    \subsection{Drug Repurposing Targets}
\begin{figure}[H]       
      \centering  
      \includegraphics[width=\textwidth]{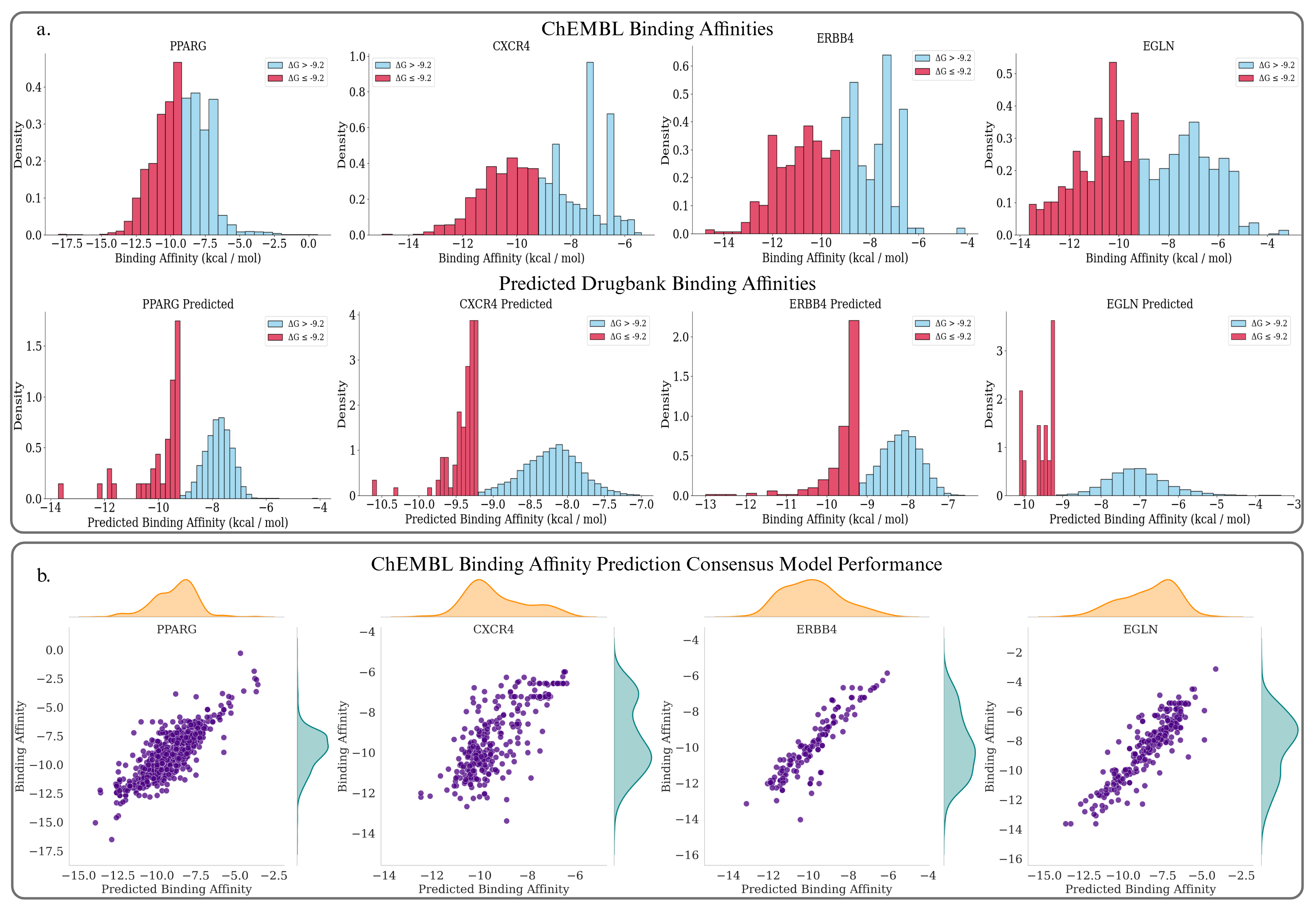}
      \caption{a. (Top) Distributions for ground truth binding affinity scores for each target taken from ChEMBL. (Bottom) Distributions for predicted binding affinity scores for DrugBank small compounds. Predictions were made from a consensus of two Gradient Boosted Decision Tree Regressors trained on a Transformer based and an ECFP based molecular representation of the ChEMBL binders. Colors are split by an effective binder threshold of -9.2 $\Delta G$.  b. Plots demonstrating the performance of the prediction models on ChEMBL testing data for each target. }
      \label{fig:binding}     
    \end{figure}

    In this section, we pivot to computationally evaluating drug repurposing candidates for our selected targets. Having assessed the therapeutic potential of our topologically significant genes, as well as the availability of public binding affinity data, we settled on focusing the remainder of our analysis on the following genes / pathways (See Figure \ref{fig:binding}). 
    
    \subsubsection{Inhibiting ERBB4}
    In Table \ref{tab:erbb4}, we list the top 10 predicted binders for ERBB4. Ibrutinib, developed primarily as a BTK inhibitor, exhibits nanomolar affinity for ERBB4 and has been shown to suppress cell growth and MEK and ERK signaling in cancer cell lines \cite{Rauf2018}. Afatinib irreversibly inhibits EGFR, HER2, and ERBB4, thus blocking ErbB receptor phosphorylation and downstream signaling \cite{Hirsh2015}. Similarly, zanubrutinib and acalabrutinib have demonstrated off-target activity against the ErbB family and may modulate ERBB4 signaling \cite{Dostalova2023}. Varlitinib is a reversible pan-HER inhibitor with nanomolar potency against ERBB4 and has shown efficacy in HER overexpressing biliary tract cancers \cite{Tan2019}. Canertinib is an irreversible pan-ErbB tyrosine kinase inhibitor with confirmed activity against ERBB4 \cite{Ayati2020}. AEE-788 is a dual EGFR/HER2 and VEGFR inhibitor developed to suppress tumor growth and angiogenesis; it also inhibits ERBB4 (HER4) with an IC$_{50}$ of 0.16~$\mu$M, suggesting potential modulation of ERBB4 signaling \cite{Traxler2004}. PD-168393 is a covalent pan-ErbB inhibitor that potently targets ERBB4 and induces autophagy without apoptosis in MPNST cells, but enhances cytotoxicity when combined with lysosomal stress \cite{Kohli2012}. HKI-357 has shown potent inhibition of HER2 in vitro and in vivo, with sustained suppression of receptor phosphorylation \cite{Tsou2005}.

\begin{table}[htbp]
    \centering    
    \caption{Top‐ranking DrugBank compounds and their predicted binding affinities for ERBB4}
    \label{tab:erbb4}    \begin{tabular}{llcc}
        \toprule
        DrugBank ID & Drug Name & \textbf{Pred.\,$\Delta G$ (kcal\,mol$^{-1}$)} &Included In Training Set\\
        \midrule
        DB09053& Ibrutinib
& -13.00& Yes\\
        DB08916& Afatinib
& -12.55&Yes\\
 DB15035& Zanubrutinib
& -12.41&Yes\\
 DB05944& Varlitinib
& -11.77&Yes\\
 DB12558& AEE-788
& -11.74&Yes\\
 DB08564& & -11.48&No\\
 DB13002& HKI-357
& -11.39&No\\
 DB05424& Canertinib
& -11.37&Yes\\
 DB07662& PD-168393
& -11.28&Yes\\
        DB11703&  Acalabrutinib& -10.98&Yes\\
        \bottomrule
    \end{tabular}

\end{table}

In addition to the highest-scoring binders based solely on affinity prediction, we further identified thirteen ERBB4-targeting compounds that also satisfied CNS-relevant criteria, including strong predicted binding affinity ($\Delta G_{\mathrm{pred}} < -9.2$ kcal/mol) and high BBB permeability and probability of P-gp inhibition and substrate interaction. Representative compounds are shown below; the full set is provided in Table \ref{tab:erbb42}. DB12016 (Ponesimod), an antagonist of sphingosine-1-phosphate receptor 1 (S1PR1), has shown potential in AD models by modulating microglial activity, alleviating neuroinflammation, and enhancing amyloid$\beta$ clearance \cite{Zhu2023}. DB11805 (Saracatinib), is a Src/Abl kinase inhibitor for cancer therapy, has shown promise in AD by targeting Fyn kinase and improving cognitive outcomes at lower doses than used in cancer therapy \cite{Ramos2024}. DB01222 (Budesonide) is a synthetic, inhaled glucocorticoid that mitigates asthma-induced neuroinflammation, which may help reduce neuronal loss and suggests a neuroprotective potential \cite{Xia2014}. DB15489 (Mexazolam) is an anxiolytic oxazolo-benzodiazepine whose active metabolite, chlornordiazepam, exhibits a pharmacodynamic profile that supports strong anti-anxiety effects with minimal sedation compared to other benzodiazepines \cite{Fernandes2022}. DB05410 (NCX 1022) is a nitric oxide-releasing derivative of hydrocortisone that modulates early skin inflammation by regulating leukocyte recruitment, resulting in faster and more effective anti-inflammatory responses than hydrocortisone \cite{Hyun2004}. DB07249 is an anilinoquinazoline inhibitor that targets c-Src tyrosine kinase and exhibits strong anti-tumor activity in a rat xenograft model based on 3T3 cells transformed with Src \cite{Ple2004}. DB00248 (Cabergoline), a dopamine receptor agonist, improved memory and reduced A$\beta$ 42 and p-tau levels in an AD model. It exerted neuroprotective and autophagy-enhancing effects via AKT/mTOR, GLT-1/P38-MAPK, and ERK1/2 pathways \cite{Joodi2025}.  DB09295 (Talniflumate) is an anti-inflammatory drug that exerts neuroprotective effects by inhibiting astrocytic ASCT2–NLRP3 interaction, thereby reducing neuroinflammation and preserving dopaminergic neurons in Parkinson’s disease models \cite{Liu2023}. 

\begin{table}[htbp]
    \centering    
    \caption{Repurposing Candidates Targeting ERBB4 with Favorable CNS Properties and Strong Binding Affinities}
    \label{tab:erbb42}
    \resizebox{\textwidth}{!}{%
    \begin{tabular}{llclllc}
        \toprule
        DrugBank ID & Drug Name & \textbf{Pred.\,$\Delta G$ (kcal\,mol$^{-1}$)} & BBB & pgp\textsubscript{sub} & pgp\textsubscript{inh}  &Included In Training Set\\
        \midrule
        DB12016& Ponesimod
& -9.62& 0.95& 0.00& 0.14&No\\
        DB13643& Loprazolam
& -9.61& 1.00& 0.00& 0.07&No\\
 DB11805& Saracatinib
& -9.49& 0.96& 0.00&0.45&No\\
 DB01222& Budesonide
& -9.49& 1.00& 0.02&0.05&No\\
 DB15489& Mexazolam
& -9.34& 1.00& 0.04&0.45&No\\
 DB05410& NCX 1022& -9.33& 0.81& 0.00&0.04&No\\
 DB07117& & -9.31& 1.00& 0.01&0.00&No\\
 DB07249& & -9.31& 0.96& 0.00&0.04&No\\
 DB00248& Cabergoline
& -9.29& 1.00& 0.00& 0.01&No\\
 DB09295& Talniflumate
& -9.29& 0.96& 0.00& 0.00&No\\
 DB09383& Meprednisone
& -9.28& 0.82& 0.00& 0.46&No\\
 DB13208& Prednylidene
& -9.25& 0.79& 0.01&0.05&No\\
        DB12549&  Pyrazoloacridine& -9.21& 0.99& 0.07&  0.15&No\\
        \bottomrule
    \end{tabular}
    }
    \vspace{0.5em}
    \parbox{\textwidth}{
        \small
    CNS-relevant ADMET properties of top-ranking DrugBank compounds predicted to bind ERBB4. All selected compounds satisfied binding affinity thresholds ($\Delta G_{pred}<-9.2$ kcal/mol), with high predicted BBB permeability ($>$70 \%) and low probabilities of acting as Pgp substrates or inhibitors ($<$50 \%).
        }
\end{table}

    \subsubsection{PPARG Agonists}
    
    In Table \ref{tab:pparg} we list the top 10 predicted binders for PPARG. The drug Efatutazone is a potent PPARG agonist of the thiazolidinedione class. While it has been explored in clinical trials for anticancer therapies, this drug has not been applied to PPARG agonism in the context of Alzheimers Disease, representing a missed opportunity \cite{Kato2012_Efatutazone}. MK-0533 is a selective partial PPARG modulator, which are being explored in clinical trials as safer options for neuro-degeneration as they preserve anti-inflammatory gene programs with fewer metabolic liabilities \cite{Cheng2019_PPARReview}. Lobeglitazone is a dual PPARA and PPARG agonist. In previous studies on AD affected mice, it lowered cortical A$\beta$ plaques, shifted microglia to an anti-inflammatory phenotype, and improved motor function\cite{Lee2021}. Like Efatutazone- Rivoglitazone and Edaglitazone are also TZD class PPARG agonists. They have been proposed for AD trials by researchers and experiments with Rivoglitazone have shown promising results regarding its anti-inflammatory signature in vitro. For Edaglitazone, however, clinical data is still pending \cite{Hong2018, Yamada2013}. Aleglitazar, Muraglitazar, Imiglitazar are also dual PPARA / PPARG inhibitors. Several mouse studies have demonstrated their promise in promoting an anti-inflammatory shift \cite{Herzig2019, Fuentes2014}. 
    
\begin{table}[htbp]
    \centering    
    \caption{Top‐ranking DrugBank compounds and their predicted binding affinities for PPARG}
    \label{tab:pparg}    \begin{tabular}{llcc}
        \toprule
        DrugBank ID & Drug Name & \textbf{Pred.\,$\Delta G$ (kcal\,mol$^{-1}$)} &Included In Training Set\\
        \midrule
        DB11894& Efatutazone& -13.72&Yes\\
        DB15242& MK-0533& -12.17&Yes\\
 DB09198& Lobeglitazone& -11.91&Yes\\
 DB07863& & -11.77&Yes\\
 DB12511& Imiglitazar& -11.72&Yes\\
 DB09200& Rivoglitazone& -10.71&Yes\\
 DB08915& Aleglitazar& -10.47&Yes\\
 DB06519& Edaglitazone& -10.40&No\\
 DB06510& Muraglitazar& -10.28&Yes\\
        DB01941&  LG-100268& -10.25&Yes\\ \bottomrule
    \end{tabular}

\end{table}

Beyond the top 10 PPARG binders, an additional 11 compounds were selected based on moderate binding affinities ($\Delta G_{\mathrm{pred}} < -8.5$ kcal/mol) and CNS-relevant ADMET filetering. A summary of these candidates is presented below. DB07242 is a $\beta$-carboline derivative; related analogs have demonstrated inhibitory activity against kinases such as PIM1, CLK, DAPK3, and BIKE \cite{Huber2012}.  DB00248 (Cabergoline) is a dopamine D2 receptor agonist approved for the treatment of hyperprolactinemia, and has been reported to normalize prolactin levels in most patients. \cite{Verhelst1999}.  Preclinical studies suggest that Cabergoline, either in its original form or as active metabolites, can penetrate the BBB and reach central targets, making it a promising candidate for CNS drug repurposing \cite{StrolinBenedetti1990}. DB01218 (halofantrine) is a lipophilic antimalarial agent effective against chloroquine resistant strains \cite{Cenni1995}. While some reports have suggested a high volume of distribution indicative of potential CNS penetration, inconsistent pharmacokinetic findings leave its BBB permeability uncertain. DB15367 (LY-2623091) is a nonsteroidal mineralocorticoid receptor antagonist with stable pharmacokinetics observed in phase 1 and 2 trials, showing consistent exposure regardless of disease status \cite{Wang2017}. DB11794 (Berzosertib) is a selective ATR kinase inhibitor. In a phase I clinical trial, it demonstrated tolerable safety and preliminary efficacy when combined with gemcitabine in patients with refractory solid tumors \cite{Middleton2021}. DB01611 (Hydroxychloroquine) is an aminoquinoline derivative originally developed as an antimalarial and later repurposed for anti inflammatory purposes \cite{VanGool2001}. It has been evaluated in an AD clinical trial, suggesting potential relevance to neuroinflammation and CNS accessibility.

\begin{table}[H]
    \centering    
    \caption{Repurposing Candidates Targeting PPARG with Favorable CNS Properties and Strong Binding Affinities}
    \label{tab:pparg2}
    \resizebox{\textwidth}{!}{%
    \begin{tabular}{llclllc}
        \toprule
        DrugBank ID  &Drug Name& \textbf{Pred.\,$\Delta G$ (kcal\,mol$^{-1}$)} & BBB & pgp\textsubscript{sub} & pgp\textsubscript{inh}  &Included In Training Set\\
        \midrule
        DB07242&& -9.17& 1.00& 0.02& 0.39&No\\
        DB00248&Cabergoline& -9.17& 1.00& 0.00& 0.01&No\\
 DB01218& Halofantrine& -9.09& 0.85& 0.04& 0.15&No\\
 DB02132& Zenarestat& -9.07& 0.72& 0.00& 0.00&No\\
 DB15367& LY-2623091& -8.99& 1.00& 0.01& 0.21&No\\
 DB07270& & -8.95& 0.71& 0.00& 0.16&No\\
 DB11794& Berzosertib& -8.95& 0.86& 0.00& 0.04&No\\
 DB13937& LGD-3303& -8.91& 0.90& 0.00& 0.02&No\\
 DB07218& & -8.88& 0.94& 0.01& 0.28&No\\
 DB01611& Hydroxychloroquine& -8.87& 0.87& 0.18& 0.31&No\\
        DB14970&Alobresib& -8.82& 0.99& 0.00& 0.00&No\\
        \bottomrule
    \end{tabular}
    }
    \vspace{0.5em}
    \parbox{\textwidth}{
        \small
    CNS-relevant ADMET properties of top-ranking DrugBank compounds predicted to bind PPARG. All selected compounds satisfied binding affinity thresholds ($\Delta G_{pred}<-8.5$ kcal/mol), with high predicted BBB permeability ($>$70 \%) and low probabilities of acting as Pgp substrates or inhibitors ($<$50 \%).
}
\end{table}

    \subsubsection{Inhibiting CXCR4}
    
    In Table \ref{tab:cxcr4} we list the top 10 predicted binders for CXCR4. The top predicted drug is Mavorixafor, which is a known oral CXCR4 antagonist used to treat WHIM syndrome. It has recently been explored for controlling chronic inflammation in neurodegeneration \cite{mavorixafor}.  MSX-122 is a partial CXCR4 antagonist which has been tested in pre-clinical stroke models and demonstrated reduced neutrophil infiltration \cite{GHASEMI2022108863}. Burixaofr is also known to block CXCR4 and lower CXCL12 driven trafficking, with clinical trials underway for modulating inflammation in WHIM syndrome \cite{Schaer2019}. It's applications to neurodegeneration have not yet been explored. Avotaciclib and Mocetinostat are CDK and HDAC inhibitors with no known interactions with CXCR4, but which are known to attenuate pro-inflammatory signalling cascades that drive glial activation and secondary neuronal injury.

\begin{table}[htbp]
    \centering    
    \caption{Top‐ranking DrugBank compounds and their predicted binding for CXCR4}
    \label{tab:cxcr4}
    \begin{tabular}{llcc}
        \toprule
        DrugBank ID & Drug Name & \textbf{Pred.\,$\Delta G$ (kcal\,mol$^{-1}$)} &Included In Training Set\\
        \midrule
        DB05501& Mavorixafor& -10.63&Yes\\
        DB12715& MSX-122& -10.61&No\\
 DB07809& & -10.29&No\\
 DB16652& Avotaciclib& -9.84&No\\
 DB11970& Burixafor& -9.74&No\\
 DB12076& Surotomycin& -9.72&No\\
 DB08441& & -9.71&No\\
 DB02022& & -9.70&No\\
 DB11830& Mocetinostat& -9.70&No\\
        DB12027&  Serdemetan& -9.68&No\\ \bottomrule
    \end{tabular}

\end{table}
 Additionally, our CNS-relevant ADMET screening for CXCR4 resulted in 11 compounds satisfying the stringent binding affinity criterion ($\Delta G_{pred} < –9.2$ kcal/mol) . Five illustrative candidates are highlighted below; the complete results are available in Table \ref{tab:cxcr42}. DB02022 (Toxopyrimidine) is a pyrimidine-based pyridoxamine antagonist that induces seizures via GAD inhibition and GABA depletion \cite{Rindi1959}. DB12740 (CC-115) is a dual mTOR/DNA-PK inhibitor under clinical investigation for cancer therapy \cite{Munster2019}. DB13069 (Nimustine) is a nitrosourea alkylating agent clinically evaluated via convection-enhanced delivery in children with diffuse intrinsic pontine glioma (DIPG) in a multicenter phase II trial, demonstrating therapeutic potential \cite{Saito2025}. DB12177 (Eplivanserin) is a highly selective $5\text{-}\mathrm{HT}_2$ receptor antagonist developed for the treatment of insomnia \cite{RinaldiCarmona1992}.  DB16954 (Ezeprogind) is a small molecule drug developed to treat Progressive Supranuclear Palsy (PSP), a tauopathy-related neurodegenerative disorder, by targeting the Progranulin–Prosaposin (PGRN–PSAP) axis and enhancing lysosomal function \cite{Verwaerde2025}.

\begin{table}[H]
    \centering    
    \caption{Repurposing Candidates Targeting CXCR4 with Favorable CNS Properties and Strong Binding Affinities}
    \label{tab:cxcr42}
    \resizebox{\textwidth}{!}{%
    \begin{tabular}{llclllc}
        \toprule
        DrugBank ID  &Drug Name& \textbf{Pred.\,$\Delta G$ (kcal\,mol$^{-1}$)} & BBB & pgp\textsubscript{sub} & pgp\textsubscript{inh}  &Included In Training Set\\
        \midrule
        DB02022&Toxopyrimidine& -9.7& 0.92& 0.19& 0.00&No\\
        DB13414&Fenyramidol& -9.61& 0.83& 0.07& 0.01&No\\
 DB12740& CC-115& -9.35& 0.98& 0.02& 0.01&No\\
 DB12485& Pimonidazole& -9.34& 0.78& 0.02& 0.00&No\\
 DB13069& Nimustine& -9.3& 0.96& 0.02& 0.01&No\\
 DB12522& Toreforant& -9.28& 0.98& 0.00& 0.10&No\\
 DB07244& & -9.28& 0.94& 0.08& 0.10&No\\
 DB12177& Eplivanserin& -9.26& 0.76& 0.00& 0.10&No\\
 DB08707& & -9.24& 1.00& 0.00& 0.13&No\\
 DB16954& Ezeprogind& -9.24& 1.00& 0.20& 0.18&No\\
        DB15091&Upadacitinib& -9.23& 0.79& 0.14& 0.18&No\\
        \bottomrule
    \end{tabular}
    }
    \vspace{0.5em}
    \parbox{\textwidth}{
        \small
        CNS-relevant ADMET properties of top-ranking DrugBank compounds predicted to bind CXCR4. All selected compounds exhibited strong binding affinity thresholds ($\Delta G_{pred}  < –9.2$ kcal/mol), with high predicted BBB permeability ($>$70 \%) and low probabilities of acting as Pgp substrates or inhibitors ($<$50 \%).
}
\end{table}

    \subsubsection{Inhibiting EGLN}
    
    In Table \ref{tab:egln} we list the top 10 predicted binders for EGLN. Several of those listed are known to exhibit HIF-stabilising effects that can lessen hypoxic damage particularly in renal-anaemia in CKD. Desidustat and Molidustat are both EGLN inhibitors that have demonstrated HIF stabilisation along with elevation of brain VEGF and tight junciton proteins in mouse models \cite{JOHARAPURKAR2022100102}. Daprodustat is an FDA approved EGLN oral inhibitor. ITI-214, Aplaviroc, Gusacitinib and Dalpiciclib are not known to interact with EGLN, but they all modulate pathways that either improve cerebral blood-flow, curb leukocyte recruitment or reduce microglial cytokine production \cite{MartinBlondel2016}. For instance, Aplaviroc is a known CCR5 chemokine-receptor agonist. CCR5 blockade dampens leukocyte trafficking into the CNS and has been neuro-protective in stroke and Alzheimer models \cite{Joy2019}. N-(R-Carboxy-Ethyl)-$\alpha$-(S)-(2-Phenylethyl) exhibited the highest predicted binding affinity, but has not yet been investigated to our knowledge for its impact on hypoxia related neuroprotection, warranting additional research. 
    
\begin{table}[htbp]
    \centering    
    \caption{Top‐ranking DrugBank compounds and their predicted binding affinities for EGLN}
    \label{tab:egln}    \begin{tabular}{llcc}
        \toprule
        DrugBank ID & Drug Name & \textbf{Pred.\,$\Delta G$ (kcal\,mol$^{-1}$)} &Included In Training Set\\
        \midrule
        DB02505& & -10.13&No\\
        DB16135& Desidustat & -10.09&No\\
 DB15642& Molidustat & -10.04&Yes\\
 DB15039& ITI-214 & -9.63&No\\
 DB06497& Aplaviroc & -9.59&No\\
 DB11682& Daprodustat & -10.02&Yes\\
 DB15670& Gusacitinib & -9.54&No\\
 DB17456& Dalpiciclib & -9.47&No\\
 DB12690& LY-2584702 & -9.46&No\\
        DB16080&  Acolbifene & -9.37&No\\
        \bottomrule
    \end{tabular}

\end{table}

Complementing our top binders for EGLN, we identified three structurally diverse compounds fulfilling both the binding affinity requirement ($\Delta G_{\mathrm{pred}} < -8.5$ kcal/mol) and CNS-related pharmacokinetic benchmarks (Table \ref{tab:egln2}). DB08149, a pyrrolo[2,3-d]pyrimidine-based small molecule, is originally developed as an ATP-competitive kinase inhibitor with demonstrated PKB$\beta$ and PKA inhibition as demonstrated in enzymatic essays \cite{Caldwell2008}. DB07606, a pyrazolo[3,4-d]pyrimidinone compound, was initially designed to target CDK pathways and has shown strong enzymatic and cellular activity in preclinical studies \cite{Markwalder2004}.  DB07227 (L‑778,123) is a dual farnesyltransferase and geranylgeranyltransferase-I inhibitor that has been evaluated in phase I clinical trials for oncology indications, including combination therapy with radiotherapy in pancreatic cancer \cite{Martin2004}. 
    
\begin{table}
    \centering    
    \caption{Repurposing Candidates Targeting EGLN with Favorable CNS Properties and Strong Binding Affinities}
    \label{tab:egln2}
    \resizebox{\textwidth}{!}{%
    \begin{tabular}{llclllc}
        \toprule
        DrugBank ID  &Drug Name& \textbf{Pred.\,$\Delta G$ (kcal\,mol$^{-1}$)} & BBB & pgp\textsubscript{sub} & pgp\textsubscript{inh}  &Included In Training Set\\
        \midrule
        DB08149  && -8.81& 1.00& 0.11& 0.45&No\\
        DB07606  && -8.73& 0.89& 0.02& 0.03&No\\
        DB07227  &L-778123& -8.72& 0.98& 0.03& 0.34&No\\
        \bottomrule
    \end{tabular}
    }
    \vspace{0.5em}
    \parbox{\textwidth}{
        \small
        CNS-relevant ADMET properties of top-ranking DrugBank compounds predicted to bind EGLN. All selected compounds satisfied binding affinity thresholds ($\Delta G_{pred}  < –8.5$ kcal/mol), with high predicted BBB permeability ($>$70 \%) and low probabilities of acting as Pgp substrates or inhibitors ($<$50 \%).
}
\end{table}

    \section{Discussion}
    \subsection{Motivation of the Sheaf Laplacian for Network Analysis}
    It should be noted that successful application of topological methods such as persistent homology and persistent Laplacians hinges on the integration of important non-spatial biological information. We can try to build a filtration from selected atoms or use generalized distance to encode biochemical interactions. These tricks are included in the approach called element-specific persistent homology (ESPH)\cite{cang2017topologynet}. 
    In the cellular (co)sheaf framework, restriction or extension maps can be understood as interactions between cliques, and a cleverly designed sheaf can integrate our prior knowledge about the network into the persistent (co)sheaf (co)homology and sheaf Laplacians. For example, by labeling each gene by their respective up or down regulation according to Log2FC. On the other hand, sheaves can also be constructed in a naive and formal manner when the question is more about distinguishing different networks where the underlying topological information is the same. An elementary situation is when we have two types of nodes $A$ and $B$ in the network. Different distributions of $A$ and $B$ in the network might lead to different network behaviours or functions. One can construct multiple sheaves by assigning two numerical quantities to $A$ and $B$ and the spectra of Laplacians will be different for different distributions. The cellular (co)sheaf framework can be applied jointly with previous approaches such as element-specific persistent homology to integrate more non-spatial biological information and achieve better results. 
    
    \subsection{Review of Potential Repurposing Candidates}
    Across our four targets we collected a total of 707, 1469, and 1299 inhibitors labeled by IC50 values for ERBB4, CXCR4, and EGLN respectively, and 4933 agonists labeled by EC50 for PPARG from ChEMBL and PubChem to train our Machine Learning based binding affinity prediction models. After deploying each model on 8865 Drugbank small compounds, we identified 276, 104, 15, and 38 potentially suitable binders for each respective target. In Tables \ref{tab:erbb4},  \ref{tab:pparg},  \ref{tab:cxcr4} and \ref{tab:egln}, we list each top 10 prediction and note that among these drugs several are already known as effective binders and in trial to determine their potential neuroprotective properties. Notably, some of the top-ranked predictions were not part of the original training set but are already known as agonists or antagonists for the respective targets. For example, Desidustat (ZYAN1), an EGLN inhibitor that has completed Phase 2 clinical trials for the treatment of anemia in CKD \cite{Parmar2019}. Similar target-specific compounds, including known binders to ERBB4 and PPARG, also appeared among the top predictions. This observation suggests that our molecular fingerprint–based binding affinity prediction models effectively identify compounds likely to bind each target, even when those compounds were not explicitly seen during training.
    
Following this, to screen compounds with potential as AD therapeutics while maintaining target specificity, we applied a secondary filtering step that incorporated both predicted binding affinity and predicted CNS-relevant ADMET properties, BBB permeability and P-glycoprotein (Pgp) inhibition/substrate probabilities, critical determinants of CNS drug efficacy and retention.  This step enabled us to find candidates with both strong target engagement and favorable CNS drug-like profiles. Other ADMET parameters such as CYP inhibition or volume of distribution were excluded to avoid excessive filtering and to preserve chemical diversity—not only to identify viable repurposing candidates, but also to leave open the possibility of using these for future lead optimization. Among the shortlisted candidates, several compounds have previously been tested in experimental models of AD, Parkinson’s disease, or neuroinflammation. For example, Ponesimod (DB12016), an antagonist of S1PR1, has shown significant promise in preclinical AD models \cite{Zhu2023}. In 5XFAD transgenic mice, it decreased A$\beta$-induced activation of microglia and astrocytes, reduced pro-inflammatory cytokines such as TNF-$\alpha$ and CXCL10, and enhanced A$\beta$ clearance through improved phagocytosis. Although S1PR1 was not among our initially selected targets, Ponesimod was identified by our pipeline based on its strong predicted binding affinity and desirable CNS relevant ADMET profile. Its inclusion highlights the ability of our method to uncover therapeutically relevant compounds that may act through alternative or convergent mechanisms relevant to AD pathology. Similarly, Hydroxychloroquine was previously evaluated in a randomized, double-blind, placebo-controlled clinical trial in early AD for 18 months \cite{VanGool2001}. Although the trial did not demonstrate significant cognitive benefit, the fact that hydroxychloroquine progressed to human testing demonstrates its prior consideration as a viable AD therapeutic, reinforcing the ability of our pipeline to recover compounds with real potential to become actual treatments. A third example is Eplivanserin (DB12177), also known as SR 46349B, a selective $5\text{-}\mathrm{HT}_2$ receptor antagonist \cite{RinaldiCarmona1992}. In a study by John P. Dougherty and Jeff Oristaglio, chronic administration of this compound enhanced memory recall in mice \cite{Dougherty2013}. Although it has not been directly evaluated in AD models, its cognitive effects suggest potential relevance for treating memory-related symptoms in AD disease and other neurodegenerative disorders. The presence of such compounds among our screening results suggests that our screening approach is capable of surfacing drug candidates that may impact Alzheimer’s-related mechanisms. Although these compounds have not yet been experimentally validated for direct interaction with the specific targets proposed in this study, future AD-related investigations may reveal their therapeutic potential. If confirmed, these findings could lead to the identification of novel mechanisms of action or support the repositioning of these compounds for new indications in AD. However, most of these studies were not designed to investigate direct interactions with the specific molecular targets highlighted in our study (ERBB4, CXCR4, PPARG, and EGLN). This presents an opportunity: compounds previously shown to affect Alzheimer’s disease–relevant processes—such as neuroinflammation, memory impairment, or synaptic dysfunction—may also exert their effects through pathways involving our selected targets. Experimental validation of such interactions could uncover new therapeutic pathways and expand the scope of drug repurposing. Notably, our prediction pipeline successfully retrieved compounds that have undergone prior evaluation in AD models or related neurodegenerative conditions, even without explicit target alignment. This reinforces the biological relevance of our computational approach and suggests that compounds without prior experimental investigation may similarly harbor therapeutic potential. All shortlisted candidates—regardless of their current validation status—could serve as valuable starting points for future development, whether through direct biochemical assays or structure-guided lead optimization.

In summary, the drug selection strategy presented in this study highlights the potential to uncover novel chemical entities with therapeutic relevance to AD.  These findings strongly support the potential for topological approaches to analyzing transcriptomics data with the goal of identifying molecular therapeutic targets, as well as Machine Learning methods for drug repurposing. These candidates warrant further investigation through in vitro and in vivo experiments to evaluate their mechanisms of action, including direct target binding. Collectively, the identified compounds offer a strong foundation for the development of next-generation AD therapeutics.

    \section{Methods}
        \begin{figure}[H]       
      \centering  
      \includegraphics[width=0.9\textwidth]{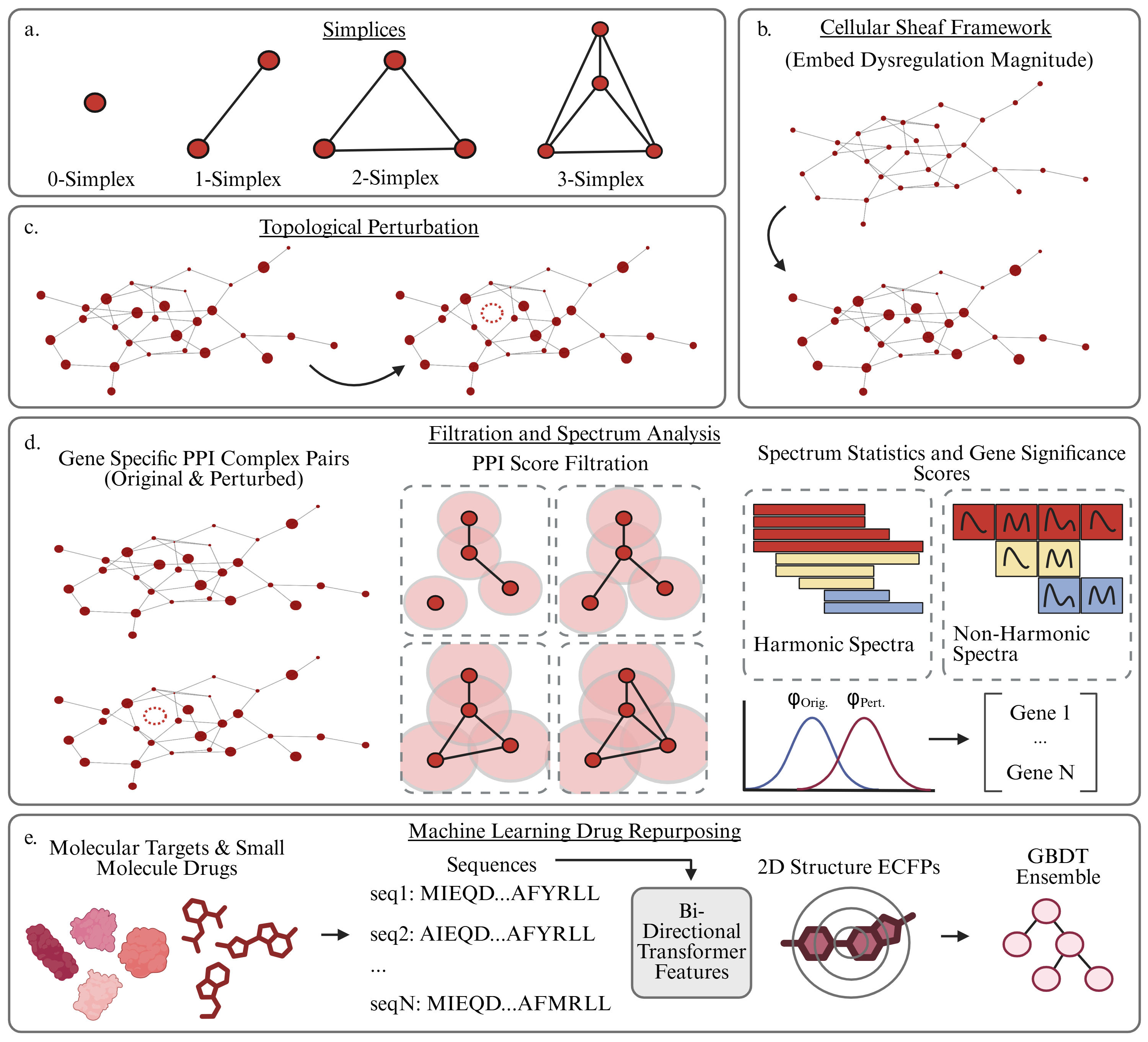}
      \caption{a. Definition of a simplex. A 0-simplex is a node, a 1-simplex is an edge connecting two nodes, and so on. b. Illustration of the Cellular Sheaf Framework. Genes in the network are labeled by the magnitude of their dysregulation, embedding non structural biological information for a richer analysis. c. Topological perturbations are performed for each gene by removing that gene from the network and analyzing the impact. d. Analysis pipeline. Topological perturbations are performed for each gene on the network. A filtration is induced via the STRING confidence scores of each P-P interaction. The spectra of the original and perturbed complexes are featurized via summary statistics, and the Wasserstein distance between the feature vectors gives the topological significance score of that gene. e. Machine Learning drug repurposing pipeline. Small molecule drugs are evaluated on each identified molecular target. Features are generated via sequence aware and 2D structure aware models, which are utilized in gradient boosted decision trees to provide binding affinity predictions. Subsequent ADMET analysis provides list of repurposeable candidates. }
      \label{fig:method}     
    \end{figure}
    \subsection{Sheaf Persistent Spectral Theory}
    We consider the PPI network as a graph with nodes representing our genes and edges indicating interactions. Constructing a simplicial complex allows for representing high-order interactions among the genes, capturing a broader range of shapes and high-dimensional relationships. Persistent Spectral Theory then reveals multiscale structures by constructing a filtration, or a nested sequence of simplicial complexes induced by a parameter representing a thresholding

    \begin{align}
        0 = K_0 \subset K_1 \subset \cdots \subset K_t = K
    \end{align}

    Here, $K$ is the largest simplicial complex in the filtration, with each $K_t$ being a sub simplicial complex at the filtration level $t$. Accompanying this filtration is a corresponding sequence of chain complexes and boundary operators at each scale 

    \begin{align}
        \left\{
\begin{array}{c}
\cdots \xrightleftharpoons[\partial_{q+2}^{*t}]{\partial_{q+2}^t} C_{q+1}^t \xrightleftharpoons[\partial_{q+1}^{*t}]{\partial_{q+1}^t} C_q^t \xrightleftharpoons[\partial_q^{*t}]{\partial_q^t} \cdots \xrightleftharpoons[\partial_1^{*t}]{\partial_1^t} C_0^t \xrightleftharpoons[]{}\begin{array}{c}
\partial_0^t \\
\partial_0^{t^*} 
\end{array}
\varnothing
\end{array}
\right\}
    \end{align}

    in which $C_q^t = C_q(K_t)$ is its chain group, $\partial_q^t : C_q(K_t) \rightarrow C_{q-1}(K_t)$ is its $q$-th boundary operator, ${\partial_q^{*t}}$is the adjoint operator of the boundary operator $\partial_q^t$, and it is relative to an inner product defined on a chain group. For each $K_t$, every $q$-simplex is oriented, and the boundary operator is applied as follows:

    \begin{align}
        \partial_q \sigma_q = \sum_{i=0}^q(-1)^i\sigma_{q-1}
    \end{align}

    Here, $\sigma_q = [v_0,...,v_q]$ is an oriented $q$-simplex within $K_t$ and $[v_0,...,\hat{v_i},...,v_q]$ represents the oriented $(q-1)$-simplex obtained by removing the $i$'th vertex from $\sigma_q$. Consider now $\mathbb{C}^{t+p}_q$ to be the subset of $C^t_q$ consisting of chains whose boundaries are in $C^{t}_{q-1}$. The $p$-persistent $q$'th boundary operator $\partial^{t+p}_q$ is then defined as a mapping from $\mathbb{C}^{t+p}_q \rightarrow C^t_{q-1}$. The $p$-persistent $q$-combinatorial Laplacian operator $\Delta^{t+p}_q$ along the filtration is then given as 

    \begin{align}
        \Delta_q^{t+p} := \partial^{t+p}_{q+1}(\partial^{t+p}_{q+1})^* + (\partial^{t}_q)^*\partial^{t}_q
    \end{align}

    If orthonormal bases are chosen for all linear spaces, then the the matrix representations of $\partial^{t+p}_{q+1}$ and $\partial^{t}_q$ can be denoted as $\mathcal{B}^{t+p}_{q+1}$ and $\mathcal{B}^{t}_q$ respectively and obtain a matrix representation of $\Delta$, denoted $\mathcal{L}^{t+p}_q$, of the form

    \begin{align}
        \mathcal{L}^{t+p}_q = \mathcal{B}^{t+p}_{q+1}(\mathcal{B}^{t+p}_{q+1})^T+ (\mathcal{B}^{t}_q)^T\mathcal{B}^{t}_q
    \end{align}

    This matrix is symmetric and positive semi-definite, ensuring that all eigenvalues are real and non-negative. The $p$-persistent $q$'th Betti numbers, representing the number of $q$-cycles persisting in $K_t$ after $p$ filtration, correspond to the nullity of $\mathcal{L}^{t+p}_q$. 

    Now, a cellular sheaf on a simplicial complex $K$ consists of the following: 
    \begin{enumerate}
        \item A simplicial complex $K$, where the face relation that $\sigma_p$ is a face of $\sigma_q$ is denoted by $\sigma_p \leq \sigma_q$
        \item An assignment to each simplex $\sigma$ of $K$ a finite dimensional vector space $\mathcal{V}(\sigma)$ and to each face relation $\sigma_p \leq \sigma_q$ a linear morphism of vector spaces denoted $\mathcal{V}_{\sigma_p \leq \sigma_q}:\mathcal{V}(\sigma_p) \rightarrow \mathcal{V}(\sigma_q)$ satisfying $\sigma_r \leq \sigma_p \leq \sigma_q \implies \mathcal{V}_{\sigma_r \leq \sigma_q} = \mathcal{V}_{\sigma_r \leq \sigma_p} \circ \mathcal{V}_{\sigma_p \leq \sigma_q}$ and $\mathcal{V}_{\sigma_p \leq \sigma_p}$ is the identity map. 
    \end{enumerate}
    The vector space $\mathcal{V}(\sigma)$ is the stalk of $\mathcal{V}$ over $\sigma$ and the linear morphism $\mathcal{V}_{\sigma_p \leq \sigma_q}$ is the restriction map of the face relation. A global section $s$ of $\mathcal{V}$ is an assignment to each simplex $\sigma$ an element $s_\sigma \in \mathcal{V}(\sigma)$ such that  $\mathcal{V}_{\sigma_p \leq \sigma_q}(s_{\sigma_p}) = s_{\sigma_q}$ for any face relation $\sigma_p \leq \sigma_q$. 
    
    Suppose then that we have a graph, or a one dimensional simplicial complex $K$ where each vertex $v_i$ is associated with a quantity $q_i \in \mathbb{R}$. Denote the edge connecting $v_i$ and $v_j$ as $e_{ij}$. We can define a sheaf $\mathcal{V}$ on $K$ such that each stalk is $\mathbb{R}$, and for the face relation $v_i \leq e_{ij}$, the morphism  $\mathcal{V}_{v_i \leq e_{ij}}$ is the multiplication by $q_j / r_{ij}$ where $r_{ij}$ is the length of $e_{ij}$. The assignment $q_i \rightarrow v_i$ and $q_iq_j/r_{ij} \rightarrow e_{ij}$ is a global section, since $\mathcal{V}_{v_i \leq e_{ij}}(q_i) = \mathcal{V}_{v_j \leq e_{ij}}(q_j) = q_iq_j/r_{ij}.$ 

    This can then be generalized to higher order simplicial complexes. Suppose we also have a nowhere zero function $F: K \rightarrow \mathbb{R}$. We can define a sheaf where each stalk is $\mathbb{R}$ and for the face relation $[v_0, v_1, ..., v_n ] \leq [v_0,v_1,..,v_n,v_{n+1}, ..., v_m ]$, the linear morphism $\mathcal{V}([v_0, v_1, ..., v_n ] \leq [v_0,v_1,..,v_n,v_{n+1}, ..., v_m ])$ is the scalar multiplication by:   
    \begin{center}
        $\frac{F([v_0, v_1, ..., v_n])q_{n+1}\cdot \cdot \cdot q_m}{F([v_0, v_1, ..., v_n,v_{n+1}, ..., v_m])}$
    \end{center}

    The Sheaf Laplacian is then a combinatorial Laplacian constructed from the sheaf cochain complexes. The nullity of the $q$-th persistent sheaf Laplacian is then equal to the $q$-th $p$-persistent sheaf Betti number. The matrix representation of the persistent sheaf Laplacian is constructed similarly to the combinatorial Laplacian, with a full treatment available in the Supporting Information.
    
    \subsection{Biomarker Identification and Drug Repurposing}
    Given our PPI simplicial complex, in which the genes make up the set of 0-simplexes, the STRING PPI confidence scores facilitate the construction of an inverse rips filtration \cite{Szklarczyk2023}. We wish to now identify the topologically significant genes via topological perturbations over each scale of the filtration. We hypothesize that the genes which are most topologically significant with respect to a dysregulation scaled sheaf would correspond well to significant biomarkers of disease progression. Topological significance is calculated according to the disparity between feature vectors obtained from the original complex and a perturbed complex where the respective gene has been removed. 

    Specifically, for a clique complex $K$ and a perturbed clique complex $\hat{K}$, we obtain the sets of subcomplexes induced by filtration: $\{K_0, K_1, ..., K_p\}$ and $\{\hat{K}_0, \hat{K}_1, ..., \hat{K}_p\}$. For each step of the filtration we compute the $L_0$ and $L_1$ Persistent Sheaf Laplacians. The spectra of these Laplacians encode the topological and geometric structure of our network with respect to a logFC labeling. We remove a specific gene from the network to construct the perturbed complex. For convenience, we calculate summary statistics of the spectra: $\{\text{Min}, \text{ Mean}, \text{ Max}, \text{ Standard Deviation}, \text{ Sum}, \text{ Number of Zeros}\}$, giving us a feature vector for each subcomplex. Having obtained a feature vector for $K_i$- call it $f_i$, and $\hat{K}_i$- call it $\hat{f}_i$, we calculate the Wasserstein distance between the two: $\text{dist}(f_i,\hat{f}_i)$. The larger the distance, the more topologically significant we consider that gene to be at that scale. For robustness, we then consider the set of genes that rank among the top 25 most significant out of 200 over all scales of the filtration, giving us our final set of inferred biomarkers. 

    Given a list of biomarkers, we move to our machine learning enabled drug repurposing. We first obtain binding affinity training data for these targets from ChEMBL. These datasets are comprised of SMILES strings for the molecular compounds, each paired with some bioactivity label, specifically $\text{IC}_{50}$ for the inhibitors and $\text{EC}_{50}$ for the agonists. In addition, we retrieved small molecule drugs, categorized under either approved, investigational, or experimental status from the DrugBank database (version 5.1.12). To ensure consistency, all SMILES strings were canonicalized by the RDKit toolkit.

    In our binding affinity analysis, we employed two fingerprinting methodologies to delineate molecular structures in a format suitable for machine learning input. Specifically, we utilized a pretrained Bidirectional Transformer, which was pretained in a self-supervised manner using the ChEMBL27 dataset by Chen and colleagues \cite{Chen2021}. We also utilized a two dimensional structure based molecular fingerprinting technique called Extended-connectivity fingerprints (ECFPs) \cite{Rogers2010}. 

    We then trained two Gradient Boosting Decision Tree regressors (one for each drug representation) on the bioactivity labeled ChEMBL data, and then tested them on thousands of small compounds obtained from DrugBank. The models were finetuned using 10 fold cross validation on a 70/30 train-test split of the ChEMBL data, and after parameter tuning were refit to the entire dataset. We averaged the predicted binding affinities between the two models to obtain our final prediction. Ultimately, drugs with a predicted equilibrium dissociation / inhibitor constant (Kd / Ki) of less than 180nm (i.e. less than -9.2 kcal / mol) were deemed to be effective binders to their respective molecular target. 
    
    To align our predictions with the specific requirements of drug repurposing for AD, we applied an additional layer of filtering to the initially screened DrugBank small molecules, focusing on key CNS related physiological properties predicted by ADMETlab 3.0 \cite{Xiong2023}. As a preliminary step, we excluded mixtures or ionic compounds, identified by the presence of multiple unconnected molecular entities in their SMILES representation. We then screened for compounds with a predicted BBB penetration probability greater than 70\%, while excluding those with a Pgp inhibitor or substrate probability exceeding 50\%. Given the CNS involvement in Alzheimer’s pathology, BBB permeability is a critical requirement for therapeutic efficacy. At the same time, compounds predicted to be either P-glycoprotein substrates or inhibitors were excluded as substrates are actively transported out of the brain, reducing central availability, and inhibitors may interfere with endogenous efflux regulation, potentially leading to safety issues or altered pharmacokinetics of co-administered drugs. We retained only those molecules that satisfied both the CNS-oriented ADMET criteria and our binding affinity thresholds: –9.2 kcal/mol for CXCR4 and ERBB4, and a slightly relaxed –8.5 kcal/mol for EGLN and PPARG due to the lack of strong candidates under the stricter threshold.

\section{Conclusion}
In this study, we performed a population-level snRNA-seq analysis to identify gene targets that commonly occur in AD patients. We have demonstrated the ability of our PSL network analysis method to present novel and testable biological hypotheses as well as to confirm known biological programmes for different aspects of AD pathology. Specifically, when applied to DEG based PPI complexes from the brain vasculature, the tool identifies genes whose interaction structure suggests a previously unexplored mechanism by which protective signaling cascades in endothelial cells connect to MMP inducing cAMP / PKA machinery in pericytes, which actually drives further leakage. Additionally, when applied to the study of PPI networks constructed from DEGs in a phagocytic transition state for microglia cells, the tool identifies genes which combine to form four axes which each comprise the hallmark phenotype of disease associated microglia, confirming the reliability of our tool in uniting gene structural significance with dysregulation magnitude to recover meaningful biological programmes in differential expression analysis. In both cases the method highlighted genes which can serve as suitable therapeutic targets, hitting multiple pathology inducing pathways simultaneously. Naturally, this analysis is not limited to the study of AD, and could extend to a variety of biological networks.

Our subsequent computational drug repurposing pipeline integrated predicted binding affinity with CNS-relevant ADMET properties to identify candidate compounds targeting ERBB4, CXCR4, PPARG, and EGLN. Several shortlisted compounds have been previously tested in models of AD or other neurological disorders, providing indirect evidence supporting the validity of our pipeline. Others, though not yet explored experimentally in AD, showed favorable profiles that warrant further investigation. Together, these results showcase the potential of our approach to surface both biologically meaningful and novel repurposing candidates that could serve as entry points for experimental validation and lead optimization. Additionally, we identified 50 potential gene targets in total. While we focused our repurposing analysis on only a small subset of these genes due to their clear involvement in pathology driving pathways and availabiltity of binding affinity information, a further investigation of these other targets is certainly worthwhile. 

 Future directions of research in this area include namely the experimental validation of our hypothesized ERBB4-DLG2 signaling cascade in endothelial and pericyte cells as well as an additional application of PSL analysis to co-expression or GRN based network configurations to overcome the incompleteness of PPI relations. We would suppose that with high quality data this approach could produce additional novel insights and testable hypotheses, potentially furthering our understanding of the molecular mechanics behind Alzheimer's Disease. Also, the efficacy of drugs cannot be determined entirely in silica, and so experimental work to decide the validity of these drug repurposing recommendations would also be necessary.  

\section*{Data  Availability}

The Brain Vascular Tissue atlas is available on the Gene Expression Omnibus under accession ID GSE163577 or via the UCSC Cell-Browser Portal:  \href{https://cells-test.gi.ucsc.edu/?ds=brain-vasc-atlas&meta=Treat}{CellBrowser}. The Dynamic Cell State Evolution Data for Disease Associated Microglia cells is available on the MIT Broad Institute webpage or via the UCSC Cell-Browser Portal: \href{https://cells.ucsc.edu/?ds=rosmap-ad-aging-brain+microglia-states}{CellBrowser}. The DrugBank small compounds used for drug repurposing efforts can be found on their website: \href{https://go.drugbank.com/}{DrugBank}.

\section*{Code Availability}
The code for Persistent Sheaf Laplacian Analysis of Gene Co-expression Networks as well as a tutorial to reproduce the results stated in this study are available at the following GitHub repository: \href{https://github.com/seanfcottrell/PersistentSheafLaplacian_Network_Analysis}{PSL PPI Analysis}. The ADMET analysis was carried out using the ADMET Lab 3.0 web portal available at \href{https://admetlab3.scbdd.com/}{ADMET Lab 3.0}.

    \section*{Acknowledgments}
This work was supported in part by NIH grant  R01AI164266, National Science Foundation grants DMS2052983 and IIS-1900473,  Michigan State University Research Foundation, and  Bristol-Myers Squibb  65109.

\newpage
\bibliographystyle{plain}
\bibliography{refs}

\end{document}